\documentclass[prb,showpacs,nobibnotes,floatfix,superscriptaddress]{revtex4}

\usepackage{amsfonts,amsmath,graphicx,epsfig,latexsym}

\begin{document}

\title{Quantum Confinement and Phase Transition in PbS  nanowire }

\author{Subhasish Mandal}
\affiliation
{Department of Physics, Michigan Technological University, Houghton MI 49931}

\author{Ranjit Pati}
\affiliation
{Department of Physics, Michigan Technological University, Houghton MI 49931}

\begin{abstract}
{\footnotesize

We report first principles density functional calculations of electronic structures and energy bandgaps 
($\Delta E_g$) in  PbS nanowires 
(NW). The $\Delta E_g$ is tuned by varying the diameter of the NW - { \it  revealing the role of quantum 
confinement}.
 The compressive radial strain (CS)  on the NW  is shown to be responsible for semiconducting 
to metallic phase transition.
The conduction band (CB) of the NW, which has significant
 contribution from the excited 3{\it d}-orbital of S, is found to be more sensitive to the CS 
with the CB minimum shifting towards and eventually crossing the 
Fermi energy with increasing CS. The origin of the observed phase transition in a recent experiment is attributed
to the CS on the PbS nanowire.

}
\end{abstract}

\maketitle
\newpage
\section{INTRODUCTION}
 One dimensional nanowires have become the foremost candidates
 in building nano transistor, optoelectronic devices, nano sensors, nano electrodes and logic
 circuits \cite{logic-gate,Zhong,nano-circuit,virus,science,Markreed}. 
One of the key features that dictate their suitability in these applications is their tunable electronic
structure property and hence their energy band gap.
 In recent years, II-VI PbS hetero nanowire
 (NW) structures have drawn considerable interest for their potential applications in optical switches
 \cite{switch,photovoltic} and solar cells \cite{solarcell,solarcell1}. Controlled synthesis of PbS nanowires with diameters
 ranging from 1.2 nm to 10 nm  have been reported \cite{somo,prb,nanolett}. This presents exciting opportunities
 to explore the tunable electronic structure property of this material
 in the strong quantum confinement regime\cite{prl1,tapered,InP,Jena,Molyb,PbSe,PbSnTe,Iop}.
 PbS in the bulk phase has a cubic close packed (CCP) structure with a near infrared direct band gap of 0.41 eV at the L point
\cite{bulk}.
Current photo luminescence study reveals wide band gap for PbS nanowires due to a higher degree of
 quantum confinement \cite{somo}. Furthermore, PbS nanowires of diameter $\sim$1.2 nm grown in Na-4 Mica channels
 have been found to exhibit semiconducting to metallic phase transition at $\sim $ 300 K \cite{prb}. The thermal
 expansion mismatch at PbS and Na-4 Mica interface producing $\sim$3 GPa pressure on the nanowire, has been
 suggested as the cause for this phase transition. 

Despite these progresses made in the last few years in controlled synthesis of PbS nanowires,
no theoretical calculations have been reported on these systems to understand the quantum confinement effect
and the origin of the observed phase transition phenomenon. In this letter, we have made the first attempt to elucidate
the tunable confinement effect in PbS nanowire as well as the origin of the phase transition from  electronic
structure calculations. 
 We  have used the first principles density functional theory to study the variation of the energy band gap 
($\Delta E_g$) with the diameter of PbS nanowire. By varying the diameter(d) of the nanowire from $\sim$ 1.17
 nm to $\sim $ 3.64 nm,  the $\Delta E_g$ is found to change from 1.524 eV to 0.955 eV; this is 
substantially higher than the
 $\Delta E_g$ of $\sim$ 0.4 eV observed for the bulk PbS - {\it clearly revealing the role of the quantum confinement.} 
The reduced 
Pb-S bond length in the $\sim$1.17 nm diameter NW as compared to that in the $\sim$ 3.64 nm wire leading to
a more confined charge density
 is found to be responsible for the observed increase in
 the $\Delta E_g $ with the decrease of `d' of the wire.
The compressive radial strain (CS) is found to have a profound effect on the electronic properties of the NW.
A semiconducting to metallic phase transition is observed when the PbS nanowire is exposed to a particular CS.
 The inward radial strain changing 
 strength of the {\it p-d} hybridization  between the Pb and S atoms is accountable for the metallic behavior in
the strained nanowire. Thus we have identified that the observed phase transition is due to the 
compressive lattice strain,
 which may have been  
caused by the thermal expansion mismatch at the PbS and Na-4 Mica interface 
as suggested by the experiment \cite{prb}. 
Besides, the evolution of $\Delta E_{g}$ with radial strain offers a tantalizing route to tune the electronic and optical properties of these nanowires.
The strained nanowire is found to exhibit indirect band gap behavior while the unstrained nanowire has 
a near direct band gap property. 
The calculated radial Young's modulus and the pressure for phase transition in a representative nanowire
of d $\sim$ 1.98 nm are found to be
857 GPa and 102.8 GPa respectively. 

 The rest of the paper is organized as follows. In Sec. II, a brief description of the computaional
procedure is given. Results and discussions are presented in Sec. III followed by a brief summary in Sec. IV.

\section{COMPUTATIONAL PROCEDURE}
 As atomic level structural details of the PbS NWs are not available a {\it priori}, we used the CCP
 PbS structure as the guiding point and constructed the 1D nanowire in the observed [200]
 growth direction. We selected two layers from the bulk structure with twelve Pb atoms and twelve S atoms and
 placed them in a tetragonal
 unit cell with  guess lattice parameter $\it{c}$ along the z-axis to construct the unit cell of the NW. 
To avoid interaction between the NW and its image in the x and y- directions, we have taken a relatively large 
lattice parameter of 1.89 nm along those two directions. Subsequently, the NW structure is optimized. The 
optimum value of $\it{c}$ is found to be 0.60 nm. The same procedure is used to obtain the equilibrium NW 
structure for 
other three NWs of d $\sim$ 1.98 nm, $\sim$ 2.80 nm, and $\sim$ 3.64 nm, containing 64, 120 and 192 atoms 
in the unit cell respectively. The optimum $\it{c}$ value (0.60 nm) is found to be insensitive to the diameter of the NW. The lattice 
parameter for the PbS nanowire of diameter $\sim $ 100 nm  reported from a recent experiment \cite{pbsexpt} is  $\sim$ 0.597 nm, 
which is in excellent agreement with our calculated value of 0.60 nm.
The average distance between two Pb atoms or two S atoms
located at two corners of the optimized unit cell (shown in Fig. 1 by the dotted line) is calculated, and 
the Covalent radii
of Pb and S are added appropriately to the average distance to estimate the diameter of the NW. 
 We have used the plane wave basis functions and periodic density functional approach within the generalized
 gradient approximation (GGA) for the exchange-correlation. The valence-core interaction is
 described by the projector-augmented wave (PAW) approach. Computations are carried out using Vienna ab-initio
 simulation (VASP) code \cite{vasp}. To determine the optimized structure, we have used a $1\times 1 \times 7 $
 (4 irreducible k-points) k-point grid  within the Monkhorst-Pack (MP) scheme to sample the Brillouin zone.
 To check the accuracy and convergence of our results, we have also performed structural optimizations
 of the NW by using $1\times 1 \times 8$ and $1\times 1 \times 11 $ k-point grids within the MP scheme. The relative difference in Cohesive Energy  by 
increasing the irreducible k-points from 4 to 6 is found to be 0.001 $\% $ while the relative difference in band 
gap energy is 0.01 $\%$.  We consider the structure to be optimized when the force on an
individual atom is $\leq  0.01 $   eV/$ \AA $. The convergence criterion for the energy during the
self-consistent calculation is taken to be $10^{-6}$ eV. The  energy cut off for the plane wave
basis is  280 eV. 

\section{RESULTS AND DISCUSSIONS}
 First, we have performed  energy band structure calculations for the PbS bulk structure
 to calibrate our computational approach. We have used $ 21\times  21\times 21$ k-point grid within the MP 
sceme to sample the brillouin zone (BZ).
The energy band diagram is summarized in Fig. 2(a).
One can notice from Fig. 2(a) that the PbS in the bulk phase is a direct band gap
semiconductor. 
The optimum lattice parameter is 5.98 $\AA $. At the L-point the  $\Delta E_{g} $ is found to be 0.44 eV. 
This is in very good agreement with the 0.41 eV for $\Delta E_{g} $ 
reported from the experiment \cite{bulk}. It should be noted that the use of $18 \times 18 \times 18$ k-point 
sampling within the MP sceme to sample the BZ yields a gap of 0.42 eV at the L-point, suggesting that
convergence in $\Delta E_g$ is achieved in our calculation with respect to the number of k-points, 
adding further confidence in our results.
 As discussed in the previous section, nanowires of different diameters 
are engineered along the observed [200] growth direction and optimized.The cross sectional view of the optimized unit cell structure from a representative nanowire of diameter $\sim $ 1.98 nm is shown in Fig. 1(a).
Fig. 1(b) shows the side view of four unit cells in [200] direction. 
First, we comment on the stability. To infer the stability of the NW,
 we have calculated the cohesive energy per atom
($E_{c} $) for each NW, which is summarized in Table 1. 
One notes from Table 1 that the maximum difference in $E_{c}$
between the bulk and the NW is $\sim$0.1 eV; this suggests that the stability of the NWs are 
comparable to that of the bulk.
As expected, the difference in $E_{c}$ between PbS nanowire and bulk PbS decreases as the 
diameter of the nanowire increases. The difference in $E_{c}$ between NW of diameter $\sim$ 3.64 nm and bulk PbS
 is only $\sim$ 0.04 eV.
Second, $\Delta E_{g} $ (Table 1) is found to decrease as the diameter of the NW increases. 
For a NW of diameter $\sim$ 1.17 nm, the
energy band gap is found to be 1.524 eV, which decreases to 0.955 eV for a NW with diameter
$\sim$ 3.64 nm. 
More importantly, we found a monotonic 
decrease in the energy band gap with an increase in diameter. To develop an atomic level understanding of the observed change in
$\Delta E_{g} $, we analyzed the nearest neighbors' bond length between  Pb and S atoms within the x-y plane.
 As the diameter decreases, the average nearest neighbor distance, {\it l},  between Pb and S is found to 
decrease (Table 1),
resulting in a more confined charge density. 
 This higher degree of quantum confinement for a smaller diameter NW is resulting in an increase 
in its band gap.
\begin{figure}
\epsfig{figure=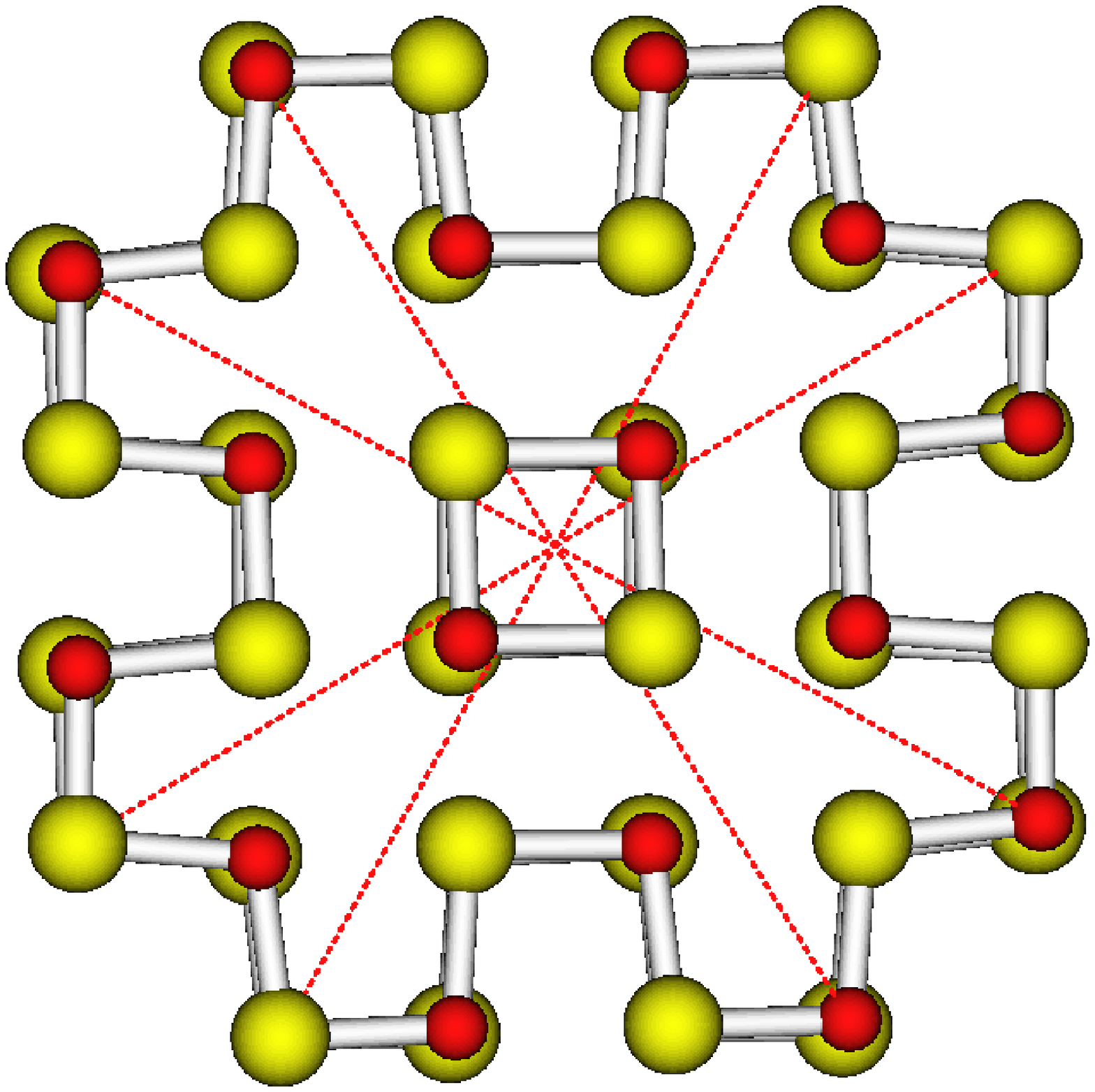, width=80pt}
\epsfig{figure=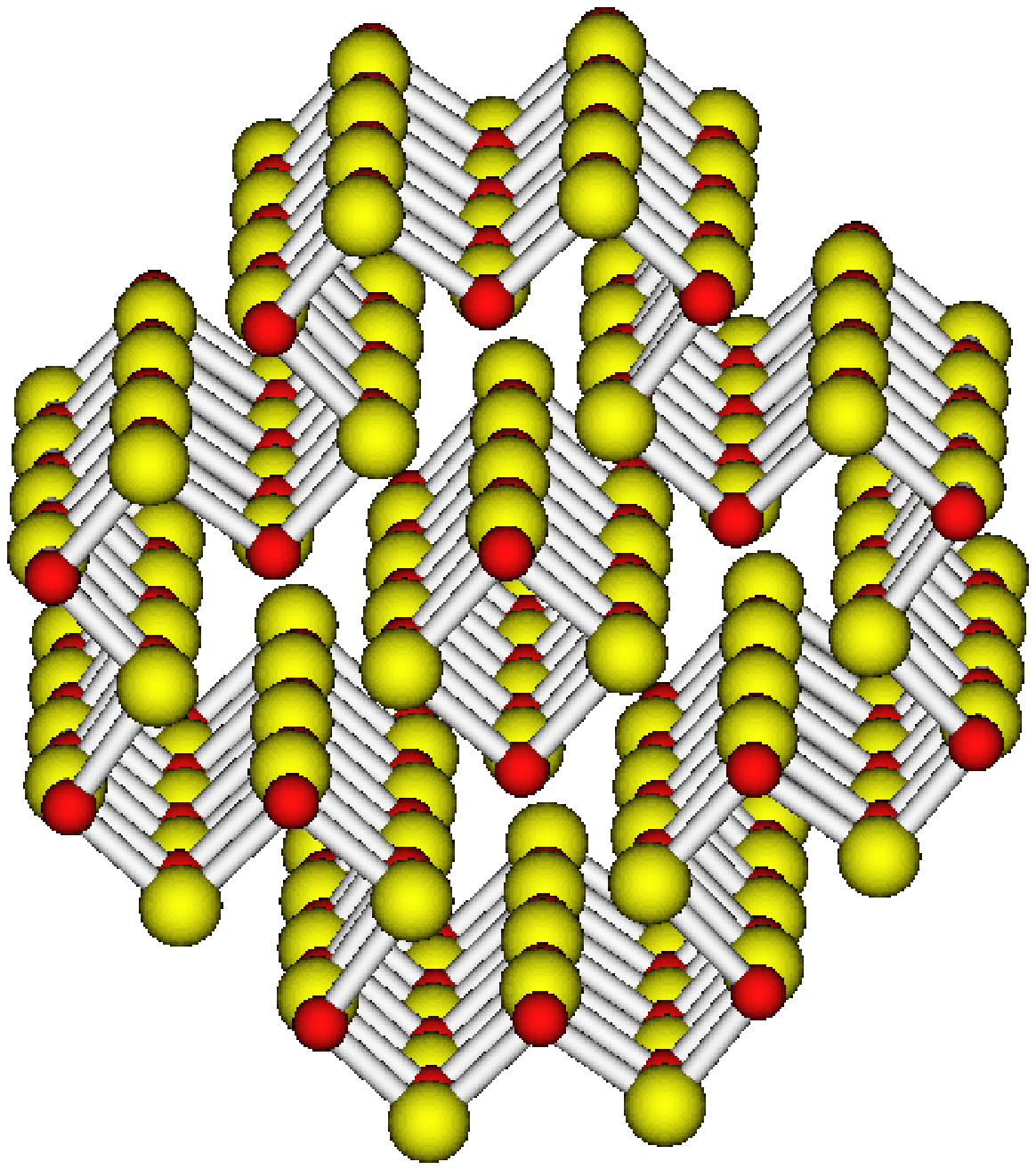, width=80pt}
\caption{(color online) Structure of PbS nanowire of d $\sim$ 1.98 nm  (a) Cross sectional view of one unit cell. (b) Side view of four unit
 cells in [200] direction. S, dark gray (red); Pb, light gray (golden). }
\end{figure}
\begin{figure}
\epsfig{figure=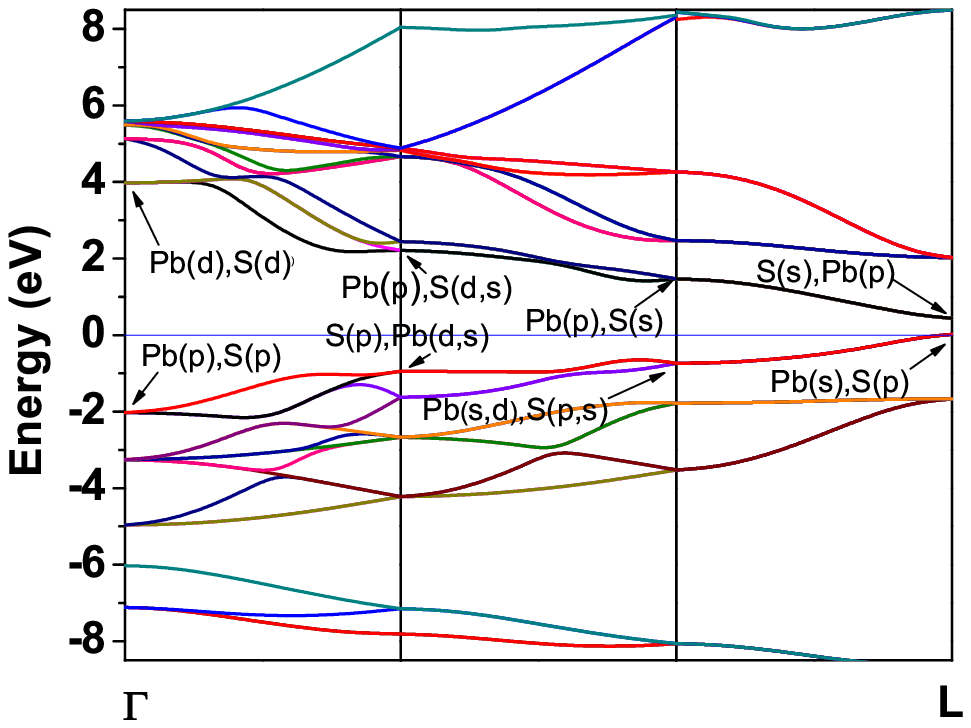,width=220pt}
\epsfig{figure=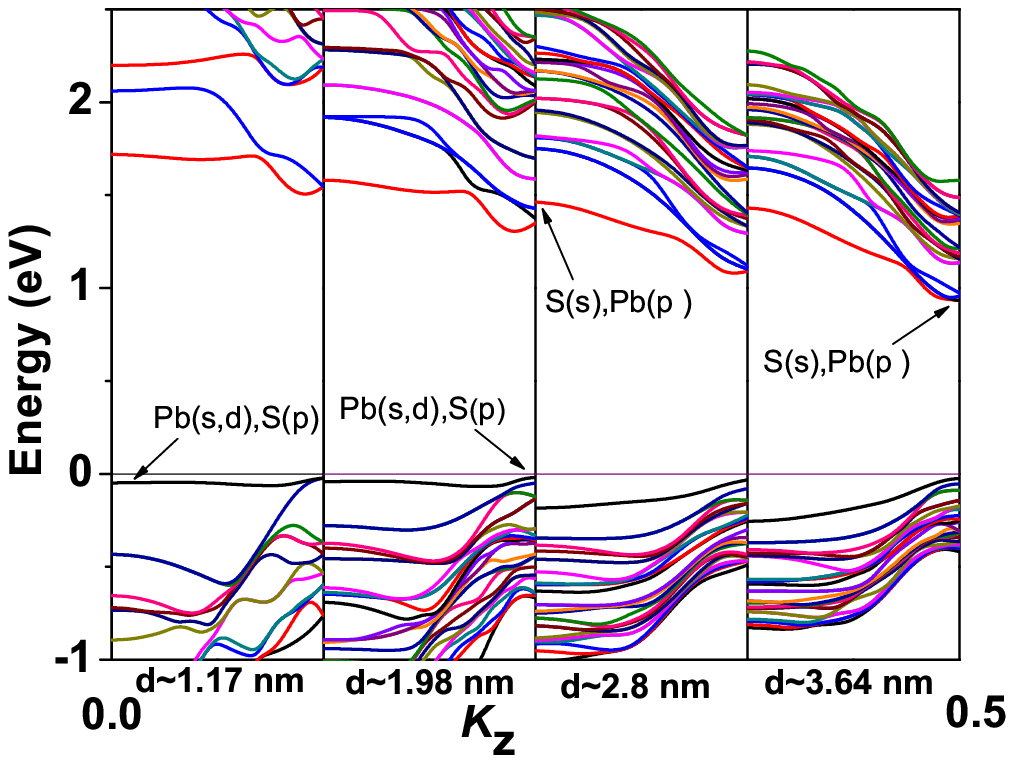,width=220pt}
\caption{ (color online) Electronic energy band structure for (a) bulk PbS; (b) PbS nanowires with
 different diameters. The Fermi level lies at {\it E=0}.}
\end{figure}

\begin{table}[h]
\caption{Calculated diameter(d), average Pb-S bond length ({\it l}), $E_c$, and $\Delta E_g $ for PbS NW. }
  \begin{tabular}{lrrr}
\hline
d(nm)\hspace*{0.5in} & {\it l} (nm)  & $E_c$ (eV)  & $\Delta E_g (eV) $ \\
\hline
1.17 &0.283 & 4.316 & 1.524 \\
1.98 &0.286 & 4.356 & 1.299 \\
2.8 &0.289 & 4.375 & 1.098 \\
3.64 &0.295 & 4.380 & 0.955 \\
Bulk & 0.299 & 4.426 & 0.44 \\
\hline
\end{tabular}
\end{table}

  To show the diameter dependence of the energy bands, we have plotted the Kohn-Sham energy bands
for all four NWs in Fig. 2(b). For larger diameter NW, the minimum energy point in the conduction band (CBM)
and the maximum energy point in the valence band (VBM) appear at the same k point. This confirms their
direct band gap property as observed in bulk PbS.
But when the diameter decreases, the CBM shifts towards the $\Gamma$-point, resulting in an indirect 
band gap behavior. Furthermore, as expected, the degeneracy in the energy bands is lifted as the diameter 
decreases. The conduction band
is found to be more sensitive to the decrease in diameter. It should be noted 
that, very recently, a similar diameter dependence feature is found in tapered silicon nanowires \cite{tapered}.
The wavefunction for the conduction
 band near the 
X-point is found to be S({\it s})-Pb({\it p}) hybridized for the NW with larger diameter. However, as the diameter decreases, the virtual 3{\it d} orbital of the S contributes to the conduction band resulting in an indirect band gap feature. 

Now we turn our discussions toward the strain induced phase transition of the nanowire. 
To replicate the effect of radial pressure 
on the nanowire, we have 
applied uniform compressive radial strain on a representative NW of diameter $\sim$1.98 nm.
The NW is allowed to relax in the [200] growth direction under radial strain. The $\%$ radial strain, $\zeta$,
 is defined as:
$\zeta =\frac{100\Delta r}{r_{0}}$, where  $\Delta r = r_{0}-r$; $r$ and $r_{0}$ are the radii 
of the nanowire with and without the radial strain respectively.
The energy band 
diagrams under different $\zeta$ are presented in Fig. 3(a). One can clearly notice 
the semiconducting 
to metallic phase transition at $\zeta= 12\%$. The conduction band, which has a contribution from the excited
3{\it d} orbital of S, is found to be more sensitive to the
compressive radial strain (CS) with the CBM shifting towards and eventually crossing the Fermi energy with
increasing CS.
The contribution from the 3{\it d} orbital of S at the CBM  develops a bonding character in the 
part of the CB wavefunction resulting in the reduction of its energy under CS.
 A similar effect of shifting the conduction band under compressive strain is noted in bulk Si \cite{strain}. 
The valence band, which is almost dispersionless, develops $\sim$ 0.4 eV dispersion width under compressive radial strain. 
We have also plotted the wavefunction for the valence and conduction bands at the $\Gamma$-point for
 three different $\zeta$s in Fig. 3(b). A distinct change in the wavefunction due to the change in the
hybridization strength between Pb and S under CS is clearly evident
 from Fig. 3(b). For the NW with d $\sim$ 1.17 nm, the semiconducting to metallic phase transition  
occurs at about $\zeta$=13 $\%$.

 To calculate the amount of pressure required for the phase transition, we have plotted the 
relative deformation
potential energy ({\it E} ) as a function of $\zeta$ in Fig. 4 for the NW of diameter $\sim$ 1.98 nm.
 The excellent parabolic behavior of 
{\it E } from Fig. 4 allows us to use 
Hooke's law to determine the radial Young's modulus, $Y_{r}$, which is defined as:
 $Y_{r}=\frac{ \delta^2 E}{\delta r^2} \frac{1}{2\pi z_{0}} $. $z_{0}$ is the length of
 the unit cell (0.6 nm) at $\zeta=0$. From the second derivative of the deformation potential {\it E},
 we have calculated the 
$\frac{\delta^2 E}{\delta \zeta ^2}$ as 1.9692 eV. Subsequently, the $\frac{ \delta^2 E}{\delta r^2}$ is obtained
 by using
the undeformed radius $r_{0}$=0.988 nm. The $Y_{r}$ is found to be 857 GPa.
 To calculate the amount of pressure ($P$) 
required for the semiconducting to metallic phase transition ($\zeta=12 \%$), we follow the simple relation 
$P=Y_{r}\frac{\Delta r}{r}$ that yields 102.8 GPa for the value of {\it P}. For the NW of d $\sim $1.17 nm,
using the same $Y_{r}$, the {\it P} is found to be 111.4 GPa.
 These values are found to be higher than 
the reported pressure ($\sim$3 GPa) for the phase transition estimated using the bulk modulus
 ($\sim$127 GPa) of PbS \cite{prb} . The substantial difference between experiment and theory can be attributed to the lower bulk modulus value \cite{prb}
used to estimate $P$ in the experiment. The use of the bulk modulus $\sim$ 127 GPa for $Y_{r}$\cite{prb} in our
 calculation would yield $P$=15.2 GPa for d$\sim$ 1.98 nm, and $P$= 16.5 GPa for d$\sim$ 1.17 nm. These are
  about five times greater than the experimentally obtained $P$. In addition, the calculation reported here is for uniform, pristine, and defect free PbS NWs, unlike the experimental case, where the PbS NWs are 
grown within the Na-4 Mica channels in different directions.

 It is well known that the energy bandgap obtained from the ground state Kohn-Sham approach does not 
represent the actual quasiparticle gap measured in the experiment. Thus, it is important to discuss whether
the many-body correlation effect, which has been found to be significant for small diameter NW, affects the 
$\Delta E_g$ obtained from GGA based DFT. It has been shown in Si-nanowire that the energy gap obtained using 
local density approximation (LDA) is significantly smaller than the observed value, which can be corrected
by using self-energy correction within the GW approximation \cite{prl3}. However, 
a recent Configuration Interaction (CI) based study in
Si nanocrystal suggests that excited state correction method does not make
notable difference as compared to the GGA based DFT \cite{Nayak}. In addition, numerous studies have also
confirmed the usefulness of DFT in predicting the trend of energy band gap in nanowires \cite{Singh,strain,modification}. 
Thus, the trends in $\Delta E_g$ that we have observed 
in PbS nanowire is not expected to change.

\begin{figure}
\epsfig{figure=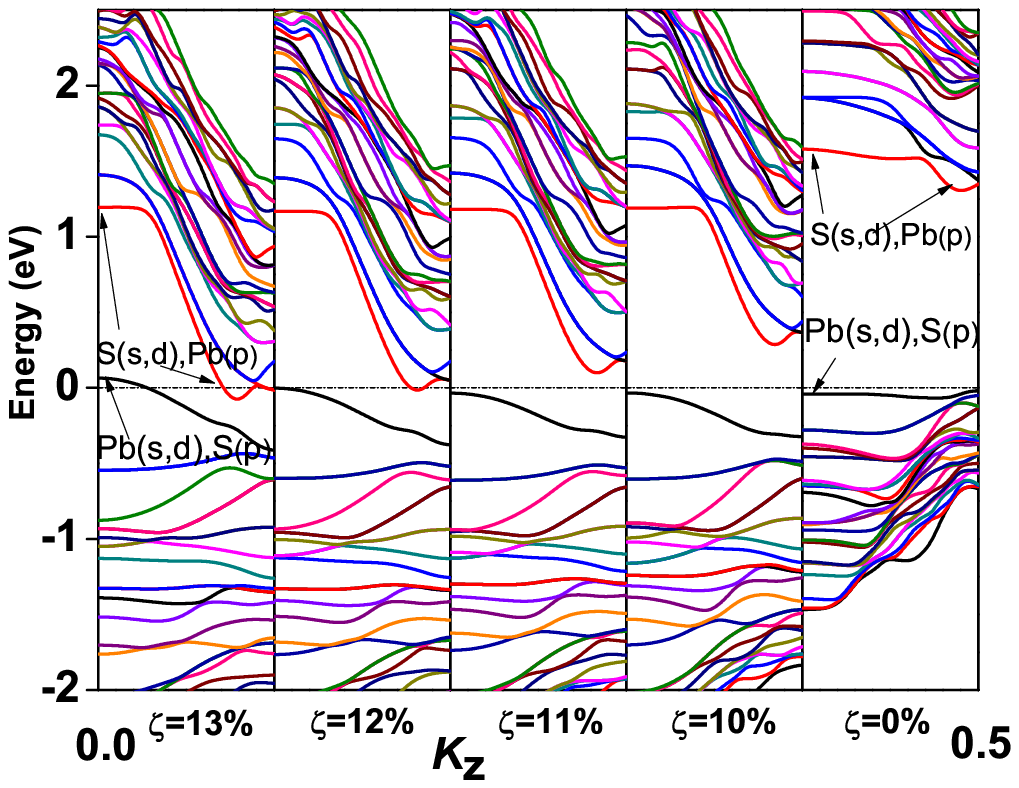, width=220pt}
\epsfig{figure=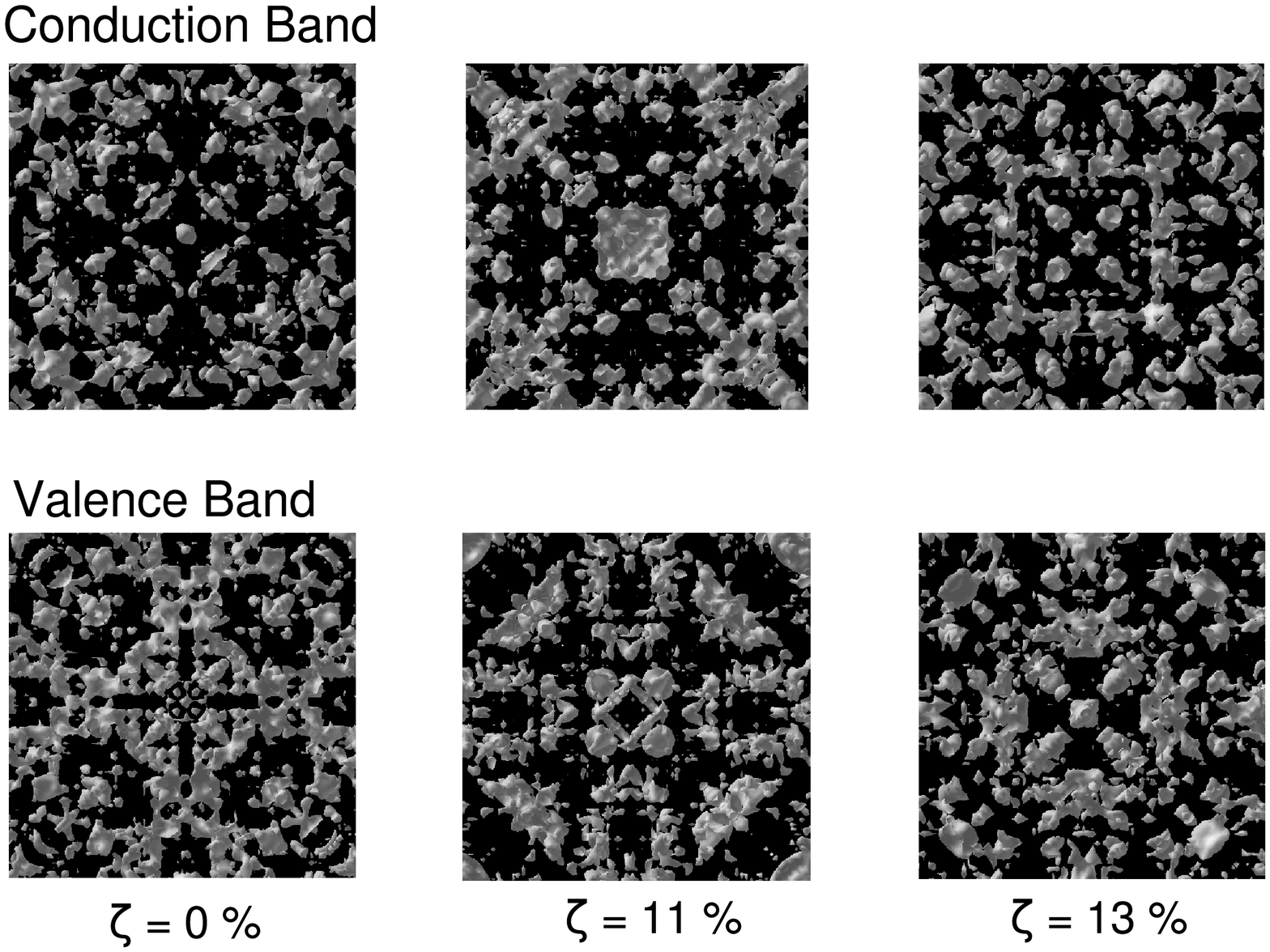, width=220pt}
\caption{(color online) (a) Electronic energy band structures for PbS nanowire with d$\sim$ 1.98 nm for 
different $\zeta$s. The Fermi level lies at E=0;(b) Cross sectional view of the electronic wavefunction
 for conduction  (above) and 
 valence (below) bands of PbS nanowire (unit cell structure is shown in Fig. 1a) with d$\sim$ 1.98 nm  at the $\Gamma$ point for three different $\zeta$s.
Isosurface value is remained fixed at 1. }
\end{figure}

\begin{figure}
\epsfig{figure=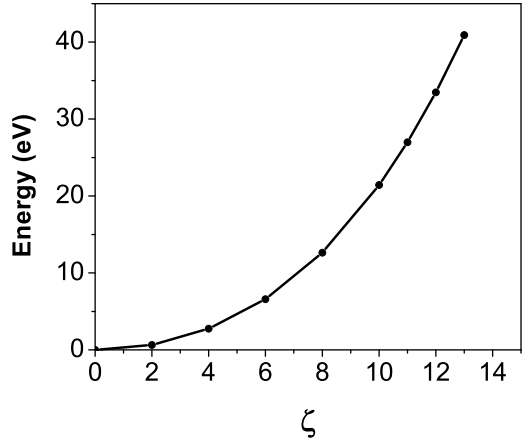, width=220pt}
\caption{ The variation of the relative deformation potential energy with $\zeta$. }
\end{figure}

\section{SUMMARY}
In summary, using the density functional approach we have probed for the first time the variation of energy band 
gap in PbS nanowire with its diameter. 
We are able to tune the $\Delta E_{g}$ of the PbS nanowire from 0.955 eV to 1.524 eV by varying the diameter
from $\sim$3.64 nm to $\sim$ 1.17 nm. This $\Delta E_{g}$ is substantially higher than the near infrared direct
band gap of 0.41 eV observed for the bulk PbS.
 The compressive radial strain on the NWs is found to have a significant effect on their electronic properties.
A semiconducting to metallic phase transition occurs at $\zeta$= 12 $\%$ for a representative NW of d $\sim$ 
1.98 nm. In addition, we have also observed the strained NW to have an indirect band gap behavior 
in contrast to the 
near direct band gap property of the NW. The conduction band of the NW, which has a significant contribution
from the excited 3{\it d}-orbital of S, is found to be more sensitive to the compressive radial strain. 
The contribution from the 3{\it d}-orbital of S at the conduction band minimum develops a bonding characteristic 
in the part of the CB wavefuntion, resulting in an energy reduction under CS with the CBM shifting
towards and eventually crossing the Fermi energy. Thus, unambiguously, we have identified that the observed phase transition
in the recent experiment is due to the CS. The tuning of the electronic structure and hence the bandgap
in PbS NWs by varying
the diameter of the NWs as well as the external strain on the NWs opens up a new route for their potential
applications in nano electronics, optical switches, and solar cells. 

\newpage
\begin{center}
{\bf ACKNOWLEDGEMENT}
\end{center}

 We thank Prof. Max Seel for very helpful discussions during this work. RP acknowledges partial support
from NSF.

\end{document}